\documentclass[twocolumn]{aastex631}
\usepackage{stfloats}
\usepackage{float}
\usepackage{graphicx}
\usepackage{natbib}
\usepackage{hyperref}
\usepackage{amsmath}
\usepackage{autobreak}
\usepackage{multirow}
\usepackage{makecell}
\usepackage{color}
\usepackage{bm}
\usepackage{threeparttable}
\usepackage[figuresright]{rotating}
\usepackage[mathscr]{euscript}
\usepackage[utf8]{inputenc}
\usepackage[T1]{fontenc}
\usepackage{amsmath}
\usepackage{makecell}
\usepackage{subfigure}
\usepackage{longtable}
\usepackage{CJKutf8}

\begin{document}
\begin{CJK*}{UTF8}{gbsn}

\title{Discovery of a variable yellow supergiant progenitor for the Type~IIb SN\,2024abfo}

\correspondingauthor{Ning-Chen Sun}
\email{sunnc@ucas.ac.cn}

\author[0000-0002-3651-0681]{Zexi Niu(牛泽茜)}
\affiliation{School of Astronomy and Space Science, University of Chinese Academy of Sciences, Beijing 100049, People's Republic of China}
\affiliation{National Astronomical Observatories, Chinese Academy of Sciences, Beijing 100101, China}

\author{Ning-Chen Sun}
\affiliation{School of Astronomy and Space Science, University of Chinese Academy of Sciences, Beijing 100049, People's Republic of China}
\affiliation{National Astronomical Observatories, Chinese Academy of Sciences, Beijing 100101, China}
\affiliation{Institute for Frontiers in Astronomy and Astrophysics, Beijing Normal University, Beijing, 102206, People's Republic of China}

\author[0000-0003-0733-7215]{Justyn R. Maund}
\affiliation{Department of Physics, Royal Holloway, University of London, Egham, TW20 0EX, United Kingdom}

\author[0000-0003-0292-4832]{Zhen Guo}
\affiliation{Instituto de F{\'i}sica y Astronom{\'i}a, Universidad de Valpara{\'i}so, ave. Gran Breta{\~n}a, 1111, Casilla 5030, Valpara{\'i}so, Chile}
\affiliation{Centre for Astrophysics Research, University of Hertfordshire, Hatfield AL10 9AB, UK}
\affiliation{Millennium Institute of Astrophysics,  Nuncio Monse{\~n}or Sotero Sanz 100, Of. 104, Providencia, Santiago,  Chile}

\author{Wenxiong Li}
\affiliation{National Astronomical Observatories, Chinese Academy of Sciences, Beijing 100101, China}

\author[0000-0001-9037-6180]{Meng Sun(孙萌)}
\affiliation{National Astronomical Observatories, Chinese Academy of Sciences, Beijing 100101, China}
\affiliation{Center for Interdisciplinary Exploration and Research in Astrophysics(CIERA), Northwestern University, 1800 Sherman Ave, Evanston, IL 60201, USA}

\author{Jifeng Liu}
\affiliation{National Astronomical Observatories, Chinese Academy of Sciences, Beijing 100101, China}
\affiliation{School of Astronomy and Space Science, University of Chinese Academy of Sciences, Beijing 100049, People's Republic of China}
\affiliation{Institute for Frontiers in Astronomy and Astrophysics, Beijing Normal University, Beijing, 102206, People's Republic of China}
\affiliation{New Cornerstone Science Laboratory, National Astronomical Observatories, Chinese Academy of Sciences, Beijing 100012, People's Republic of China}

\begin{abstract}

We report the discovery of a progenitor candidate for the Type~IIb SN\,2024abfo using multi-epoch pre-explosion images from the \textit{Hubble Space Telescope} and the \textit{Dark Energy Camera Legacy Survey}.
The progenitor exhibited a $\sim$0.7-mag decline in F814W from 2001 to 2013, followed by significant brightening and color fluctuations in the $g$, $r$ and $z$ bands. This is the first time that substantial photometric variability has been found for the progenitor of a SN~IIb. We suggest that the variability is caused by intrinsic changes in the progenitor star instead of varying obscuration by circumstellar dust. Our results show that the progenitor of SN\,2024abfo was likely a yellow supergiant star with an initial mass of 
12--18 ~$M_\odot$ for circumstellar reddening of $E(B-V)_{\rm CSM}\,<\,$0.2~mag.
Our study underscores the critical role of multi-epoch imaging surveys in revealing the final stages of core-collapse supernovae progenitors.
\end{abstract}

\section{Introduction} \label{sec:intro}

In the standard model, stars with initial masses greater than about 8~$M_{\odot}$ end their lives as core-collapse supernovae. 
Among these, Type~IIb supernovae(SNe~IIb) is a transitional class of core-collapse events, characterized by the presence of hydrogen(H) in early spectra that diminishes within tens of days \citep{1988Filippenko,2014Modjaz,2016Liu}. 
This distinctive spectral evolution implies that the progenitor of SN~IIb possessed a low-mass H envelope, suggestive of significant mass loss happening for very massive single stars \citep{2014Gal-Yam} or binary-driven stripping prior to core collapse \citep{1995Nomoto}. Based on direct detections in pre-explosion archival images, the progenitors of SNe~IIb have been confirmed to be supergiant stars(e.g., \citealp{2011Maund,2014VanDyk,2022Kilpatrick}) with effective temperatures hotter than the red supergiant progenitors for normal SNe~IIP due to the partial stripping of their H-rich envelopes. Binary companions of three SNe~IIb progenitors have also been detected \citep{2004Maund,2018Ryder,2019Maund}, which strongly supports interacting binaries as the dominant progenitor channel toward SNe~IIb.


Pre-SN stellar variability can introduce additional uncertainties in determining the initial mass and, thereby, the pre-SN evolution of the progenitor.
In recent years, substantial pre-explosion photometric variability of the progenitor stars for SNe~IIP (e.g., \citealp{2019Rui,2023Jencson,2023Soraisam,2023Niu,Xiang2024}), IIn (e.g., \citealp{2010Smith,2022Brennan}), Ibn \citep{2007Pastorello,2007Foley,2016Maund,2020Sun} and Broad-lined Ic \citep{2019Ho} has been observed.  \citet{2015Strotjohann} also reported a likely precursor for the Type~IIb SN~2012cs. 
The luminosity variations of progenitors for SNe~IIn are frequently observed \citep{2014Ofek,2021Strotjohann}, easily exceeding 1~mag; however, their specific identities and evolutionary pathways remain poorly understood, with only tentative links to very massive ($>30\,M_{\odot}$) luminous blue variables proposed \citep{2009GalYam,2017Smith}. 
Meanwhile, studies of the Type~Ibn SN\,2006jc have further demonstrated that even lower-mass stars(e.g., $<\,15$\,$M_{\odot}$) can display giant pre-SN eruptions. Progenitors of Type~IIP SNe\,2017eaw, 2023ixf, and 2024ggi also exhibited pronounced pulsational brightness variability  and may also be obscured by circumstellar dust \citep{2019VanDyk,2024Zimmerman}. 
These effects complicate progenitor interpretation--for instance, the inferred properties of SN~2023ixf’s progenitor differ markedly among studies(see Table~2 in \citealt{2024Qin}).

In this paper, we report the detection of a progenitor of the Type IIb SN\,2024abfo. 
SN\,2024abfo occurred in a nearby galaxy NGC\,1493 (z\,=\,0.003512, via the NASA/IPAC Extragalactic Database). It was first discovered by ATLAS on 2024 November 16 and was initially classified as a Type~II~SN due to the presence of a broad P-Cygni H-alpha feature in the early spectra obtained $\sim$19~hours post-discovery \citep{2024abfo_class}. 
Subsequently, \citet{Reguitti2025} reclassified it as a SN~IIb based on the disappearance of H lines in the SN spectrum obtained approximately one month after explosion.
Its redshift corresponds to a distance of 12.92~Mpc, after correcting for local velocity flows such as those induced by the Virgo cluster, and assuming a Hubble constant of $73.30\pm1.04\,{\rm km\,s^{-1}Mpc^{-1}}$ \citep{2022Riess}.
The Galactic foreground reddening towards the SN is $E(B-V)$\,=\,0.009~mag \citep{2011sfd}. Applying an extinction law of $R_V\,=\,3.1$ \citep{ebvlaw}, the Galactic extinction is $A_V\,=\,0.028$~mag. 
The extinction from the host galaxy is negligible, given that the Na\,\textsc{i}\,D absorption lines in the SN spectrum are too weak to provide a reliable estimation of reddening of the host galaxy \citep{Reguitti2025}. 
Using the reported $m_B=11.8$~mag for NGC\,1493 \citep{2022GLADE} and distance above, the galaxy luminosity–metallicity relation \citep{2004Tremonti} yields 12\,+\,log(O/H)\,$\approx$\,8.70~dex, corresponding to approximately solar metallicities \citep{2001AllendePrieto}. 
Since the SN is located 5.8~kpc from the galaxy center, we adopt a slightly subsolar metallicity of $Z=0.010$ in the subsequent analysis.

Extensive pre-explosion imaging of SN\,2024abfo are available from multiple facilities, including the \textit{Hubble Space Telescope}(HST)/Wide Field and Planetary Camera 2(WFPC2) , Dark Energy Camera Legacy Survey(DECaLS)/ Dark Energy Camera(DECam), Visible and Infrared Survey Telescope for Astronomy(VISTA)/VISTA InfraRed CAMera(VIRCAM), and the \textit{Spitzer}/Infrared Array Camera(IRAC). 
These multi-epoch, multi-band observations allow us to constrain the properties of the progenitor and investigate whether it has significant variability before explosion.
Section~\ref{sec:obsdata} describes the data and photometry procedure. The inferred physical properties of the progenitor candidate are presented in Section~\ref{sec:sed}, and we summarize our findings in Section~\ref{sec:sum}.

\begin{table*}[htbp]
\centering
\footnotesize
\caption{Observations of the SN\,2024abfo site before the explosion. \label{tab:tab1}}
\begin{tabular}{ccccc}
\hline
\hline
Instrument & Date & Filter & Exposure Time & Program ID \\
 &(yyyy-mm-dd) &(s)  & \\
\hline
WFPC2 & 2001-05-02  & F814W & 2$\times$300s & 8599$^{a}$  \\
DECam & 2013-12$\sim$2018-01 & $g$ & 7$\times$90s & 2012B$-$0001$^{b}$   \\
DECam & 2013-12$\sim$2018-01 & $r$ & 6$\times$90s & 2012B$-$0001  \\
DECam & 2013-11$\sim$2018-02 & $z$ & 6$\times$90s & 2012B$-$0001  \\
VIRCAM & 2011-12-09  & $J$ & 60s & 179.A$-$2010(E)$^{c}$  \\
VIRCAM & 2011-12-09  & $H$ & 60s & 179.A$-$2010(E) \\
VIRCAM & 2011-12-09  & $K_s$ & 60s & 179.A$-$2010(E) \\
     \hline
\end{tabular} \\
PI last name:(a) Boeker;(b) Frieman;(c) Richard.
\end{table*}

\section{Data and photometry}\label{sec:obsdata}


\subsection{SOAR Telescope}

In order to obtain the accurate position of SN\,2024abfo in the pre-explosion image, we carried out 4$\times$30s $r$-band imaging observations of the field of SN\,2024abfo on 2024 November 18 using the Goodman High Throughput Spectrograph(GHTS) equipped on the 4.1-m Southern Astrophysical Research(SOAR) telescope located at Cerro Pachon, Chile.
The combined SOAR image is shown in Figure~\ref{fig:fig1}(a), where the SN is clearly visible and remains unsaturated thanks to the short single-exposure time. 
Centers of the SN and field stars on this image were determined through two-dimensional Gaussian fitting. 
The common field stars that were detected both in the SOAR image and pre-explosion image were used to conduct the astrometric transformation. The transformation included rotation, scaling and shifting and was achieved through equation~\ref{eq1}, where (x,y)/(u,v) donates coordinates on the SOAR/pre-explosion image.
\begin{equation}\label{eq1}
\begin{bmatrix}
u \\
v
\end{bmatrix}
=
\begin{bmatrix}
P_1 & P_2 \\
Q_1 & Q_2
\end{bmatrix}
\begin{bmatrix}
x \\
y
\end{bmatrix}
+
\begin{bmatrix}
P_0 \\
Q_0
\end{bmatrix}
\end{equation}
We used least-squares minimization to derive the transformation coefficients, where iterative rejection of 3$\sigma$ outliers was employed. 
We subsequently computed transformation residuals for the common stars to estimate the uncertainty of transformation.
Detailed results of the transformed SN coordinate shall be described in the following subsections.

\subsection{Hubble Space Telescope}

 \begin{figure*} [htbp] 
    \centering %
    \includegraphics[width=\linewidth]{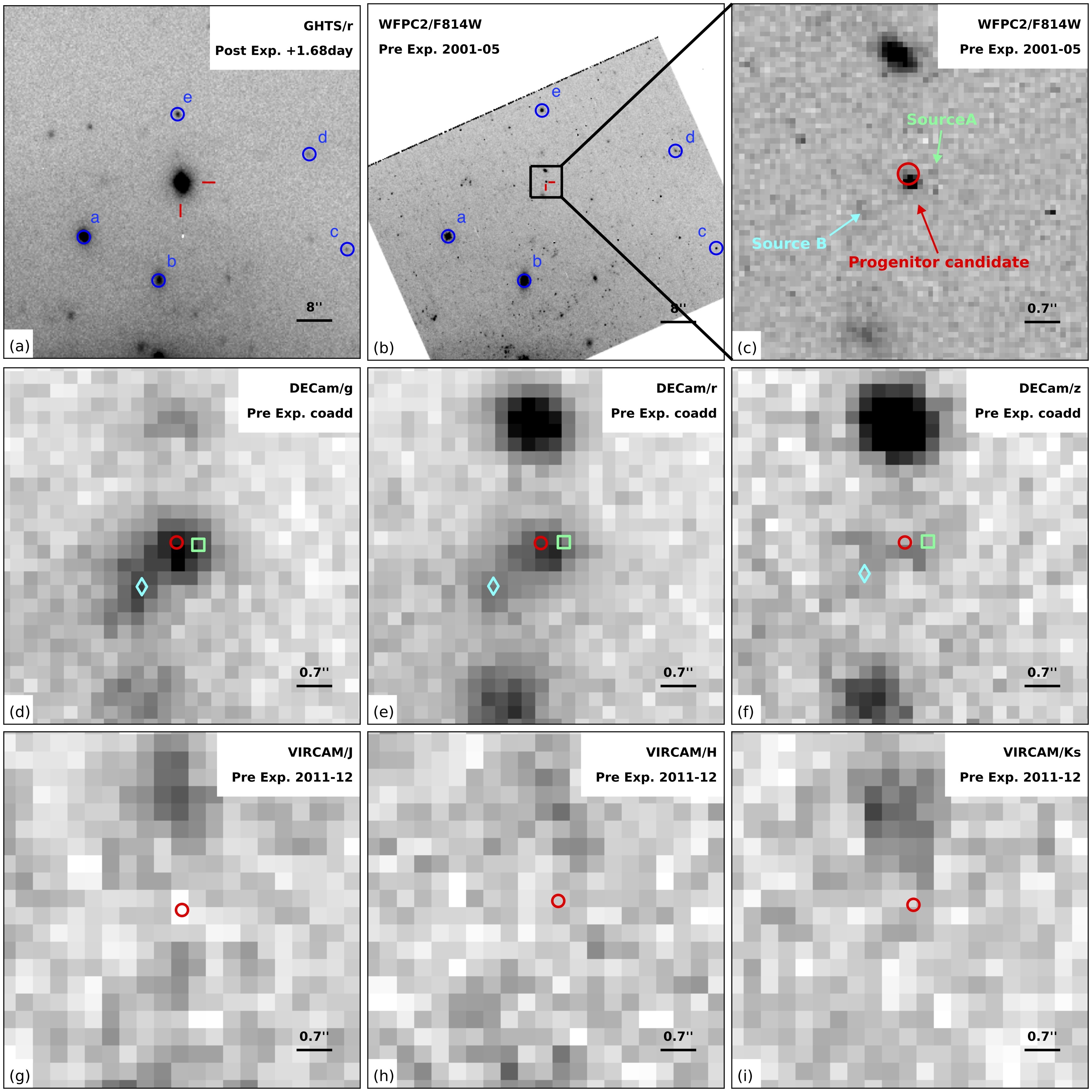}
    \caption{Pre- and post-explosion images of the site of SN\,2024abfo. 
    Each panel is labeled with the instrument, band, and epoch. All panels are aligned with north up and east to the left.
    Panel(a) and(b) have the same scale, with reference stars used for astrometric transformation between WFPC2 and SOAR circled in blue and SN site marked in red.
    The transformed SN position and its uncertainty are shown with the red circle in Panel(c). 
    Two nearby sources detected in F814W image are indicated in cyan and green, respectively.
    Panel(c) to(i) show a same zoomed-in scale. 
    About 20 common stars on SOAR and DECaLS/VISTA images are used for astrometric transformation. 
    The transformed SN position as well as two nearby sources are also indicated in the same colors as in panel(a). 
    The typical astrometric uncertainties for both DECaLS and VISTA are 0.4 pixels.
    \label{fig:fig1}}
\end{figure*}

Before the explosion, this SN site was imaged by the HST with the Wide Field Planetary Camera 2 (WFPC2) in F814W in 2001, with one 40-s exposure and two 300-s exposures (PI: Boeker). The 40-s exposure is not used in this work, and for the two 300-s exposures we retrieved the calibrated images from the Mikulski Archive for Space Telescopes\footnote{\url{https://mast.stsci.edu/search/ui/\#/hst}}. 
We employed the \textsc{iraf} task \textsc{crrej} to construct the cosmic-ray mask and combined two exposures using the \textsc{astrodrizzle} package\footnote{\url{http://drizzlepac.stsci.edu/}} for display purposes.
Point-spread function(PSF) photometry was then performed with the \textsc{dolphot} package with WFPC2 module \citep{2000Dolphin}, in which parameters were set according to recommendations in the User Guide.


When aligning the HST and SOAR frames, 5 common sources that are clearly detected in both images can be found (marked with blue circles and labelled a--e in Figure~\ref{fig:fig1}(a,b)). The common source a/b/c/d/e has $i=17.45/17.93/21.74/20.98/19.78$~mag. The transformed SN position on the WFPC2/F814W image is indicated by a red circle in Figure~\ref{fig:fig1}(c), with a radius of 2.0~HST pixels (0.2~arcsec), representing the uncertainty of the differential astrometry. This uncertainty corresponds to the mean astrometric residual of the common sources c--e, whose brightness are more comparable to the progenitor candidate (see below).

An isolated bright source is clearly visible at the transformed SN position on the F814W image with an offset of only 1.8~pixels; hereafter we call it progenitor candidate, with $m_{\rm F814W}\,=\,21.86\,\pm\,0.03$~mag(Vega). 
A faint source of $m_{\rm F814W}\,=\,24.87\,\pm\,0.27$~mag is detected 0.5~arcsec away from the progenitor candidate, as marked with a green arrow; hereafter we call it Source A. 
At this offset, the contamination from Source A to the progenitor candidate is negligible.
Another source is located 1.1~arcsec southwest of the progenitor candidate with $m_{\rm F814W}\,=\,24.630\,\pm\,0.228$~mag, indicated in a cyan arrow; hereafter we call it Source\,B. 

\begin{table}[htbp]
\centering
\footnotesize
\caption{AB-Magnitude Photometry of the progenitor candidate of SN\,2024abfo from DECaLS images. \label{tab:lc}}
\begin{tabular}{cccc}
\hline
\hline
Date & Filter & Magnitude & Error  \\
(yyyy-mm-dd) & & & \\
\hline
2013-12-09 & g & 23.13 & 0.11 \\ 
2013-12-12 & g & 22.99 & 0.18 \\
2013-12-12 & g & 23.11 & 0.18 \\
2016-09-26 & g & 23.48 & 0.18 \\
2016-12-25 & g & 22.98 & 0.16 \\
2017-12-15 & g & 22.92 & 0.18 \\
2018-01-18 & g & 22.67 & 0.17 \\
2013-12-12 & r & 22.88 & 0.16 \\
2013-12-12 & r & 22.83 & 0.16 \\
2013-12-29 & r & 22.98 & 0.14 \\
2016-09-24 & r & $>$23.05 & 0.22 \\
2017-12-15 & r & 22.86 & 0.18 \\
2018-01-18 & r & 22.76 & 0.17 \\
2013-11-14 & z & 22.98 & 0.30 \\
2013-12-21 & z & 23.13 & 0.24 \\
2013-12-22 & z & 22.99 & 0.22 \\
2016-01-18 & z & 23.19 & 0.28 \\
2016-12-18 & z & 22.62 & 0.21 \\
2018-02-18 & z & 23.28 & 0.34 \\
     \hline
\end{tabular}
\end{table}

\subsection{Dark Energy Camera Legacy Survey}\label{sec:decam}

\begin{figure}[htbp] 
    \centering %
    \includegraphics[width=\linewidth]{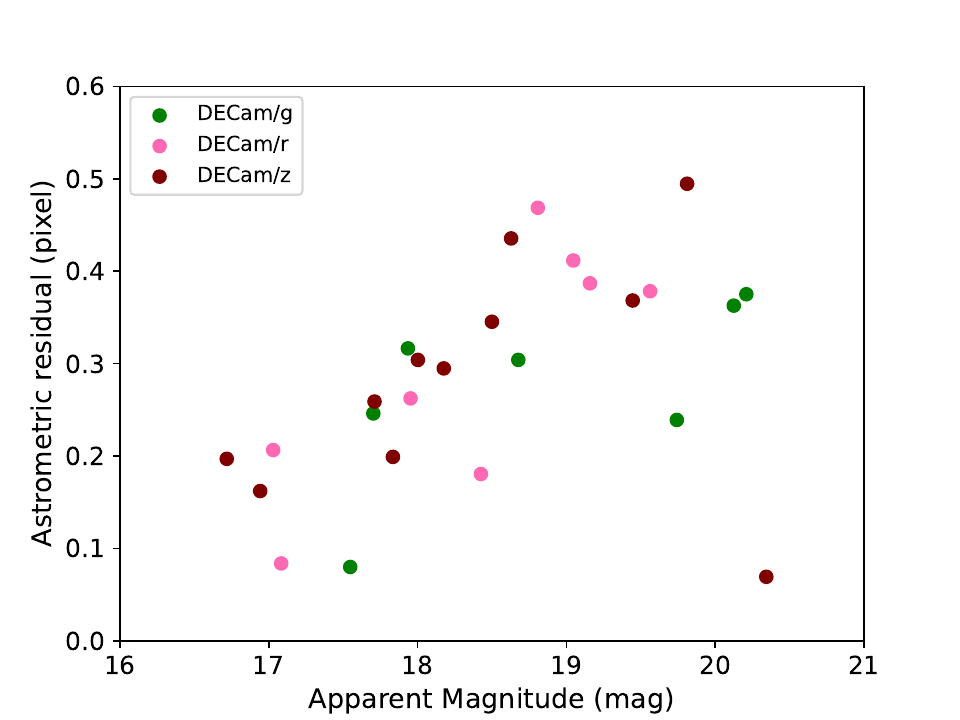}
\caption{An example of the uncertainties in differential astrometry for DECaLS images. The pixel size of DECaLS images is 0.262~arcsec.
\label{fig:ae}}
\end{figure}

The DECaLS, carried out by the 4.1-m Blanco Telescope at the Cerro Tololo Inter-American Observatory, also covered the SN site in the $g$, $r$ and $z$ bands(brick ID: 0592m462). These datasets consist of 7\,$\times$\,90-s exposures in the $g$ band and 6\,$\times$\,90-s exposures in each of the $r$ and $z$ bands from 2013 November to 2018 February. 
We retrieved the single exposure and coadd images from Data Release 9\footnote{\url{https://www.legacysurvey.org/}}. 
We selected about 10 common sources to perform the astrometric transformation between the SOAR and DECaLS images. Figure~\ref{fig:ae} shows astrometric residuals and magnitudes of the common sources. At the SN magnitude（see Figure~1 of \citealp{Reguitti2025}), the typical error of astrometric transformation is always less than 0.4 pixels (0.1~arcsec).

The transformed positions of the SN are shown in red points in Figure~\ref{fig:fig1}(d--f). The position of Source A/B is marked with a green square/cyan diamond according to its relative locations as observed in the HST image. A point-like source is visible near the SN position in the $g$-, $r$-, and $z$-band images. The SN progenitor and Source A are unresolved in the DECaLS images due to the larger pixel size of 0.262~arcsec. However, it is reasonable to assume that this point-like source is predominantly contributed by the progenitor candidate with negligible emission from Source A, since the progenitor candidate is $\sim$632 times brighter than Source A in the F814W-band images.

This source was reported in the Dark Energy Legacy Survey Data Release 1 (DR1) and Data Release 2 (DR2); the former is based on observations from 2013 August to 2016 February \citep{2018Abbott}, while the latter \citep{Dsurvey_DR2} extended the data up to 2019 January. Its aperture magnitudes (AB), \texttt{MAG$\_$AUTO}, which are measured on the coadd images, are $g=23.17\pm0.08$~mag, $r=21.89\pm0.04$~mag, $z=22.58\pm0.20$~mag in DR1 and $g=22.61\pm0.03$~mag, $r=22.22\pm0.03$~mag, $z=22.47\pm0.12$~mag in DR2.
However, the aperture model for this target, as shown by the model image, has a highly eccentric elliptical shape that overlaps with both the progenitor candidate and the nearby Source B. Consequently, the \texttt{MAG$\_$AUTO} measurements are contaminated by the Source B and not adopted in this study.
Additionally, the DECaLS performed PSF photometry on single-epoch images and reported weighted average PSF magnitudes(AB), \texttt{WAVG$_{\rm \texttt{MAG}}$PSF}, in DR2 with $g=23.05\pm0.06$~mag,  $r=22.84\pm0.06$~mag, $z=22.52\pm0.13$~mag. 
Unfortunately, PSF photometry for individual epochs is not provided.

We performed our own PSF photometry on the single-exposure images.
For each image, we selected unsaturated stars with signal-to-noise ratio exceeding 5$\sigma$ above the background and no bright stars within a radius of 35~pixels. Typically, more than 100/200/200 stars can be selected for the $g$/$r$/$z$-band images.
By using the \texttt{EPSFBuilder} of the \texttt{Photutils} package \citep{Photutils.ref}, we constructed empirical PSF models based on these selected stars. The resulting full width at half maximum (FWHM) values were in the range of 3.7–4.2 pixels, consistent with those provided by the DECaLS. 

\begin{figure*}[htbp] 
    \centering
    \includegraphics[width=0.45\textwidth]{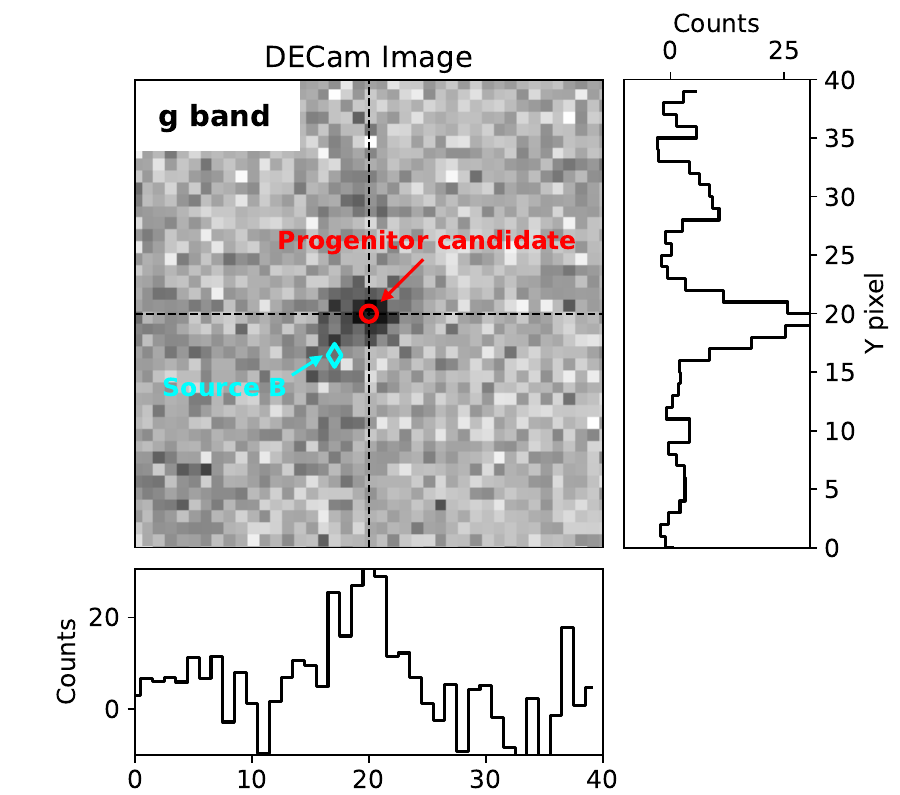}
    \includegraphics[width=0.45\textwidth]{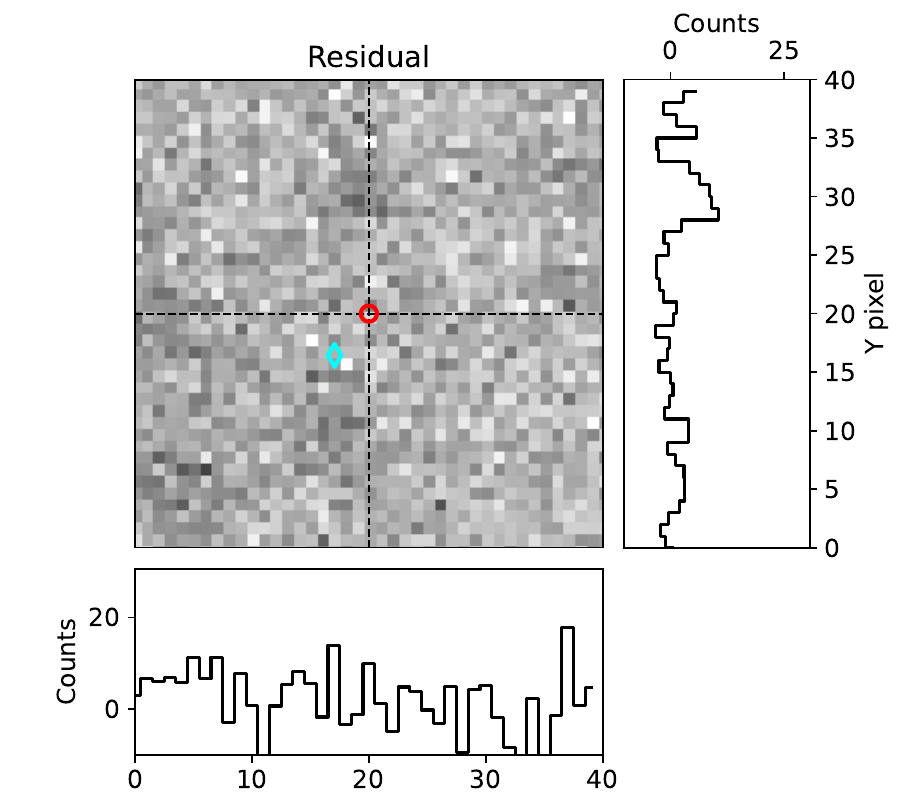}
    \includegraphics[width=0.45\textwidth]{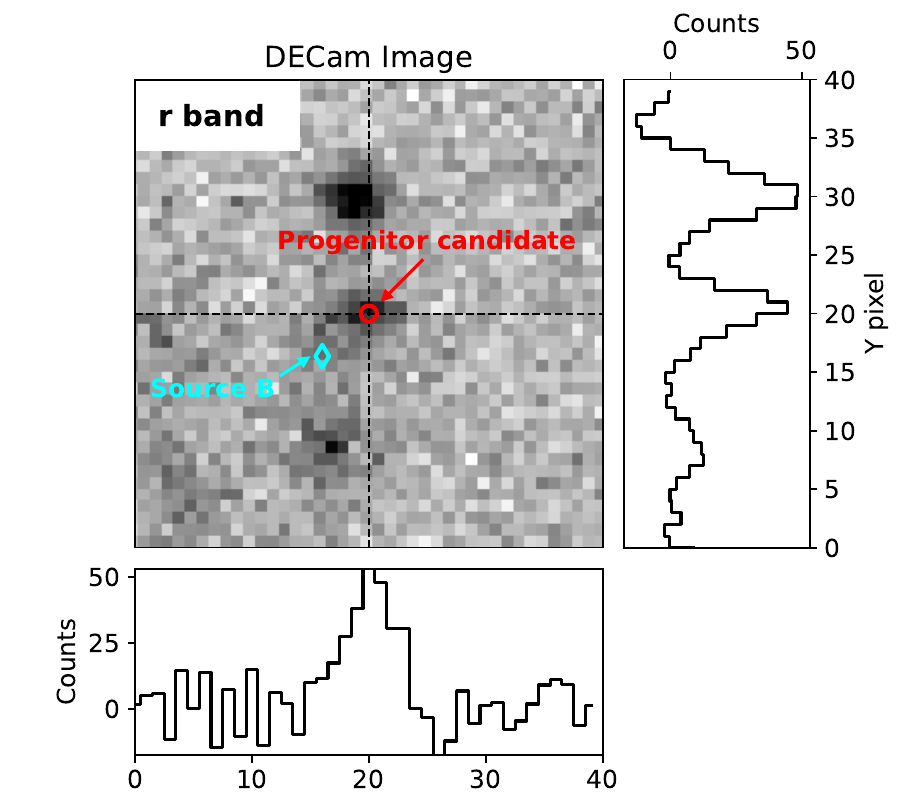}
    \includegraphics[width=0.45\textwidth]{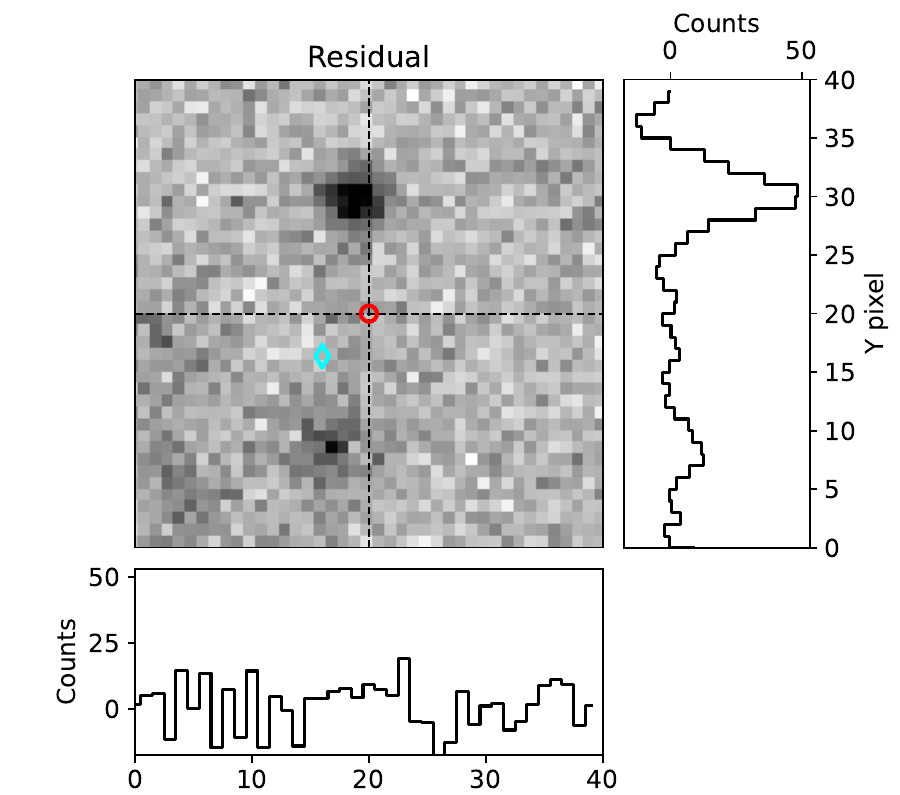}
    \includegraphics[width=0.45\textwidth]{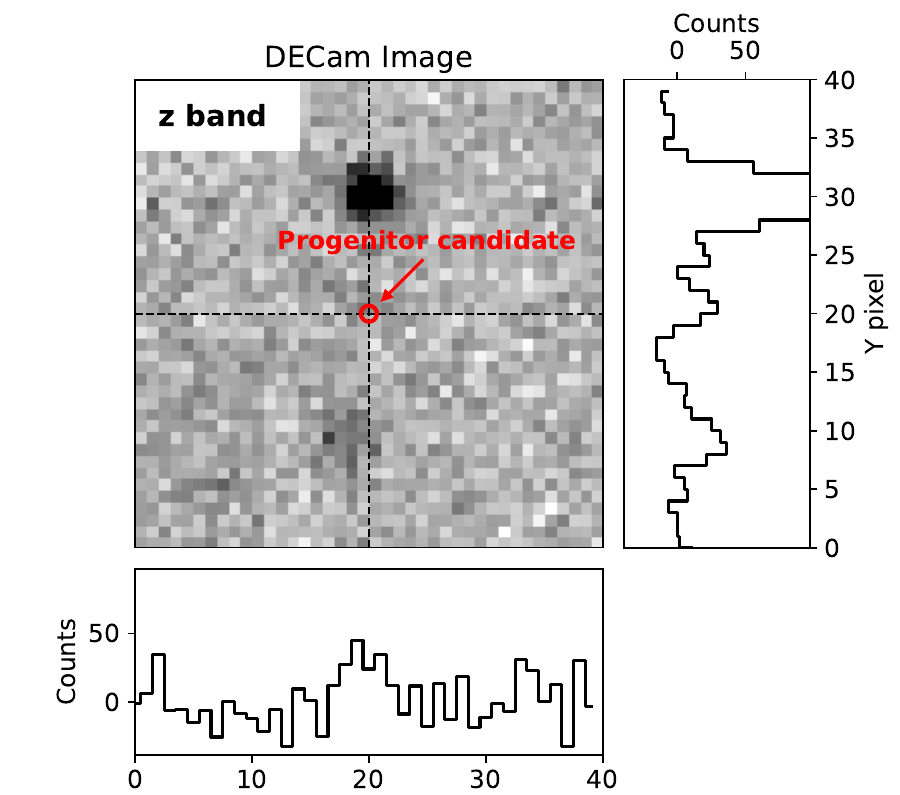}
    \includegraphics[width=0.45\textwidth]{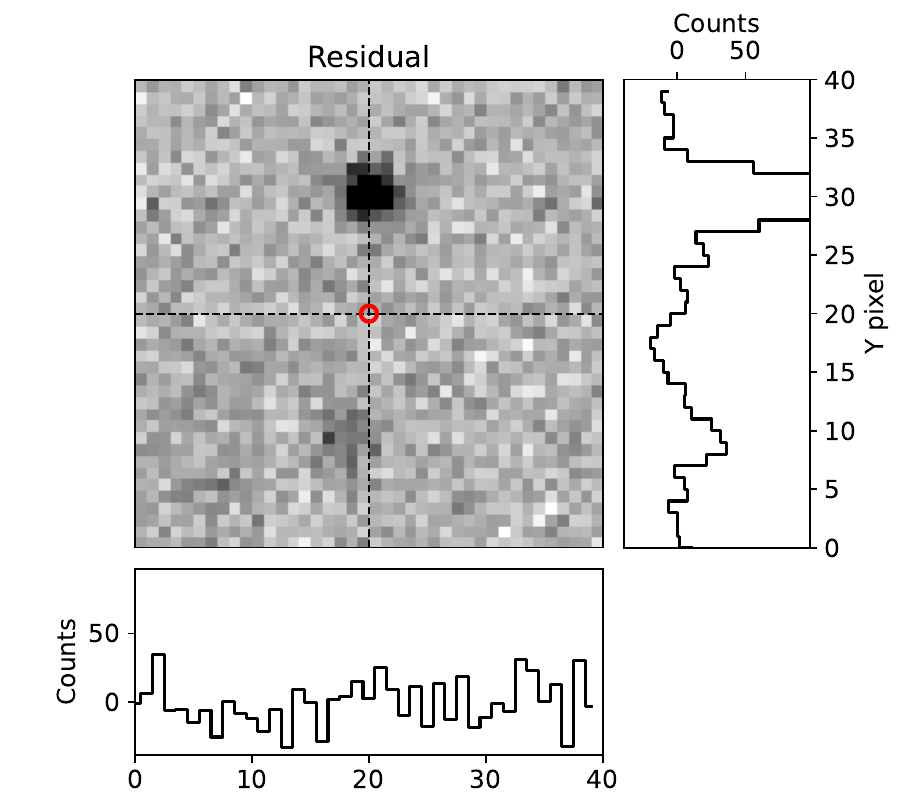}
    \caption{The PSF subtraction of images obtained in 2018 December as an example. 
    Each row shows single-exposure images before(left) and after(right) the PSF subtraction. Red and cyan scatters mark the fitted positions of the SN progenitor candidate and Source\,B, respectively.
    In the $g$ and $r$ bands, to avoid contamination from the Source\,B(cyan), we performed PSF fitting for Source\,B and the progenitor candidate simultaneously. In the z band of all epochs, Source\,B was undetected in the single-exposure image; therefore, PSF photometry was only performed for the progenitor candidate.
    Projections centered at the progenitor candidate along the x- and y-axis are also shown for each panel.
    The progenitor and Source\,B are successfully subtracted, and no visible residuals remain in the subtracted images.}
    \label{fig:gpsf}
\end{figure*}

We measured instrumental magnitudes ($m_{\rm inst}$) using the \texttt{PSFPhotometry}\footnote{\url{https://photutils.readthedocs.io/en/stable/user_guide/psf.html}} subpackge with the empirically derived PSF models. 
The progenitor candidate and nearby Source\,B are overlapped with each other in the $g$ and $r$ bands, as shown in Figure~\ref{fig:fig1}. 
For these two band, to improve the fit, the progenitor and the Source B were simultaneously fitted by assigning the same \texttt{group$\_$id} in \texttt{PSFPhotometry}. 
In all $z$-band single-exposure images, Source\,B is too faint to be detected; therefore we only made PSF fitting for the progenitor candidate.
An example of PSF subtractions is demonstrated in Figure~\ref{fig:gpsf}, where both the progenitor candidate and Source B are cleanly subtracted. 

\begin{figure}[htbp] 
    \centering %
    \includegraphics[width=\linewidth]{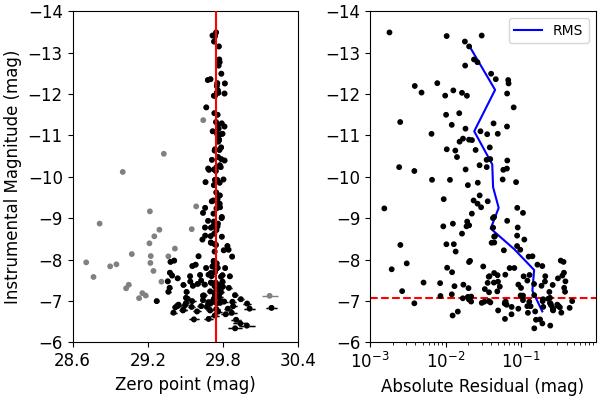}
\caption{An example of the zero point fitting for the PSF photometry of DECaLS single-exposure images. The left-hand panel plots the zero-point measurements (red vertical line) for the \textit{g} band image taken in 2018, where gray and black points represent data before and after a 3$\sigma$ clipping, respectively. 
The right-hand panel shows the distribution of absolute values of residuals, where the blue line indicates the Root Mean Square (RMS) of different instrumental magnitudes. Near the progenitor candidate's magnitude (indicated by the red dashed line), the RMS of the residuals is approximately 0.17~mag.
\label{fig:zp}}
\end{figure}

Then, we measured instrumental magnitudes for all stars in the images, and we compared their instrumental magnitudes with their PSF magnitudes from \citet{Dsurvey_DR2} in the corresponding band to determine the zero point for each image.
A few sources were discarded if their residuals within twice the FWHM radius showed significant deviations from zero. 
For the remaining stars, the zero point was calculated as ZP\,=\,\texttt{WAVG$_{\rm \texttt{MAG}}$PSF}\,$-\,m_{\rm inst}$. 
We fitted all ZPs with a Gaussian function after applying an iterative $3\sigma$ clipping. An example of the measurement and the corresponding residual are shown in Figure~\ref{fig:zp}. 
The standard deviations of the residuals around the progenitor’s instrumental magnitudes reflect the ZP uncertainties, and they are found to be much larger than the PSF fitting uncertainties. Therefore, we adopt the ZP uncertainties as an estimate of the final photometric uncertainties of the progenitor.

The derived photometry are listed in Table~\ref{tab:lc}. For the $g$ and $r$ bands, most of our single-epoch photometry agree well with the cataloged \texttt{WAVG$_{\rm \texttt{MAG}}$PSF} expect for dimming event in 2016 December and brightening event in 2018 January. 
It worth noting that the $r$-band PSF fitting of epoch in 2016 September returned a stellar centroid between the progenitor candidate and Source B, resulting in measurements of combined flux of the two sources. Therefore, we adopted this measurement as an upper limit for the brightness of the progenitor. 
We notice that the cataloged \texttt{WAVG$_{\rm \texttt{MAG}}$PSF$\_z$} is brighter than most of our $z$-band magnitudes. We suspect that this weighted average magnitude is biased by the brightening event in 2016 December. We shall discuss these variability in Section\,\ref{sec:sed}.

\subsection{Infrared Telescopes}

Figure~\ref{fig:fig1}(g--i) display pre-explosion $JHK_s$ images acquired by VISTA. Aside from a bright source located 2~arcsec north of the progenitor position and exhibiting a remarkably red color across optical-infrared wavelengths, no source was reliably detected near the SN site. The nominal 5$\sigma$ detection limits(AB) are $J\,=\,20.9$~mag, $H\,=\,20.3$~mag and $K_s\,=\,19.8$~mag \citep{2013McMahon} 

We also searched for the progenitor in \textit{Spitzer}/IRAC images. In the IRAC imaging, the SN position coincides with the wings of the northern red source, thereby preventing its detection within the infrared data.

\section{Results}\label{sec:sed}

\begin{figure*} [htbp] 
    \centering %
    \includegraphics[width=\linewidth]{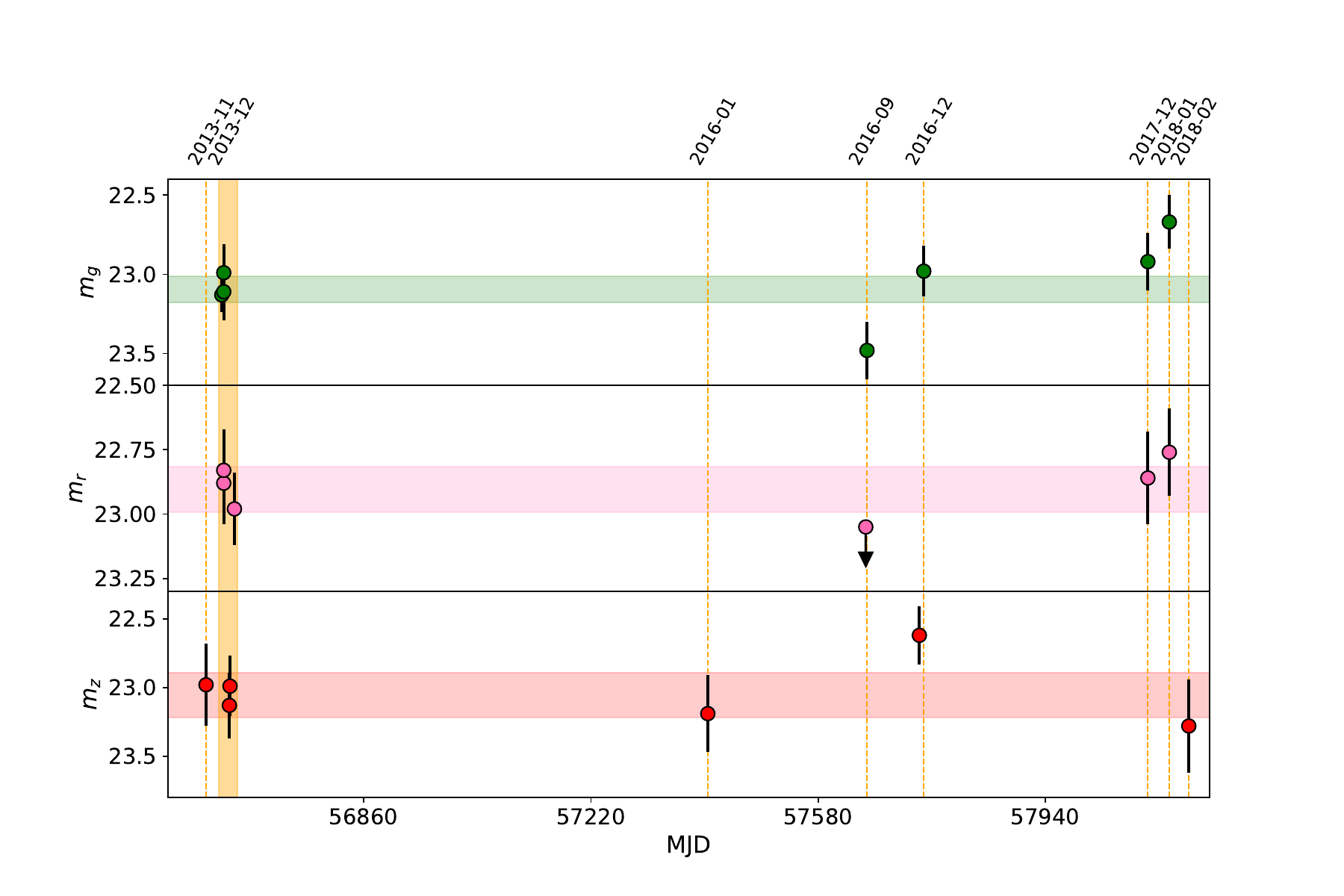}
\caption{Light curves of PSF photometry for the progenitor candidate in the $g$-, $r$-, and $z$-bands images from 2013 to 2018. The deduced explosion date of the SN is 2024 November 14 (MJD 60628) \citep{Reguitti2025}. The hatched regions denote the weighted mean values of the photometry in 2013 December and their corresponding uncertainties.
\label{fig:lc}}
\end{figure*}

Figure~\ref{fig:lc} shows the light curves of the progenitor candidate in the $g$, $r$ and $z$ bands. From 2013 November to 2013 December, there was no obvious variation for each band; so we used the weighted averaged magnitudes in 2013 December as a reference baseline as shown in Figure~\ref{fig:lc}.
Compared with that, the $g$-band magnitude decreased/increased by 0.39/0.43~mag at 2016 September/2018 January.
The $r$-band brightness contemporaneously became slightly lower/higher at those epochs. The progenitor brightened in the $z$ band by 0.43~mag in 2016 September.
Unfortunately, the sparse sampling make it difficult to accurately determine whether there is any periodicity in the variation.

\begin{figure} [htbp] 
    \centering %
    \includegraphics[width=\linewidth]{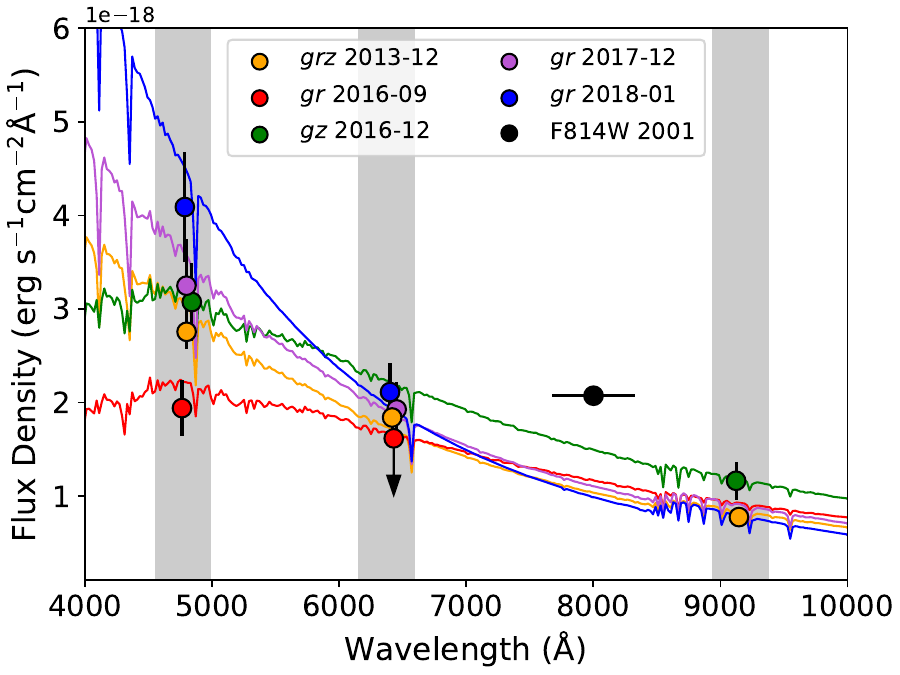}
    \caption{SEDs of the progenitor candidate at different epochs. The vertical errorbars represent 1$\sigma$ uncertainty and the grey-shaded regions indicate the bandpasses of the $g$, $r$ and $z$ bands.
    The best-fitting stellar atmospheric models with no circumstellar extinction are plotted in the same color as the data points for different epochs. 
    \label{fig:SED}}
\end{figure}

Figure~\ref{fig:SED} shows the evolution of the spectral energy distribution(SED) of the progenitor candidate. We plotted 5 key epochs where at least two bands of photometry are available. By interpolating the $g$- and $z$-band flux at 2013 December, we find the F814W band in 2001 is significantly brighter than that at 2013 December by 0.71~mag. It is also clear that the progenitor candidate significantly dimmed and possibly redwarded in 2016 September and brightened and bluewarded in 2018 January,  while the other three epoch shows mild variations. 


These features could arise from two possible scenarios. In the first scenario, the observed variations are caused by changes in the circumstellar extinction. While this scenario can explain the dimming in 2016 September by dust in the newly ejected material, it is difficult to account for the return to "normal" brightness within just three months by 2016 December as well as the sudden brightening in 2018 January and the subsequent dimming in 2018 February. 
Besides, the F814W-band magnitude decreased by 0.71~mag from 2001 to 2013 December; however, the progenitor candidate still appears quite blue in 2013 December. If the brightness decline were caused by increasing dust extinction, the progenitor in 2013 December would be extremely hot (log($T_{\rm eff}/K$)\,$>$\,4.20) and luminous (log$(L/L_{\odot})\,>\,5.97$), consistent with a very massive star with initial mass over 50--70\,$M_{\odot}$ instead of a SN~IIb progenitor. Thus, we deem this scenario very unlikely.

In another scenario, the observed variability is caused by intrinsic changes of the progenitor star. To characterize the evolution of the progenitor as it approaches the SN explosion, we performed SED fitting using the stellar atmosphere models of \citet{ck04} for the 5 key epochs with observations of at least two filters. The synthetic photometry was carried out with \textsc{pysynphot} \citep{pysynphot.ref}, and parameter optimization was performed using the \textsc{emcee} package \citep{2013emcee}. However, the possible circumstellar extinction of the progenitor remains unknown, for which we considered a range of values of $E(B-V)_{\rm CSM}$\,=\,0.0, 0.1, 0.2 and 0.3~mag.

\begin{figure*} [htbp] 
    \centering %
    \includegraphics[width=\linewidth]{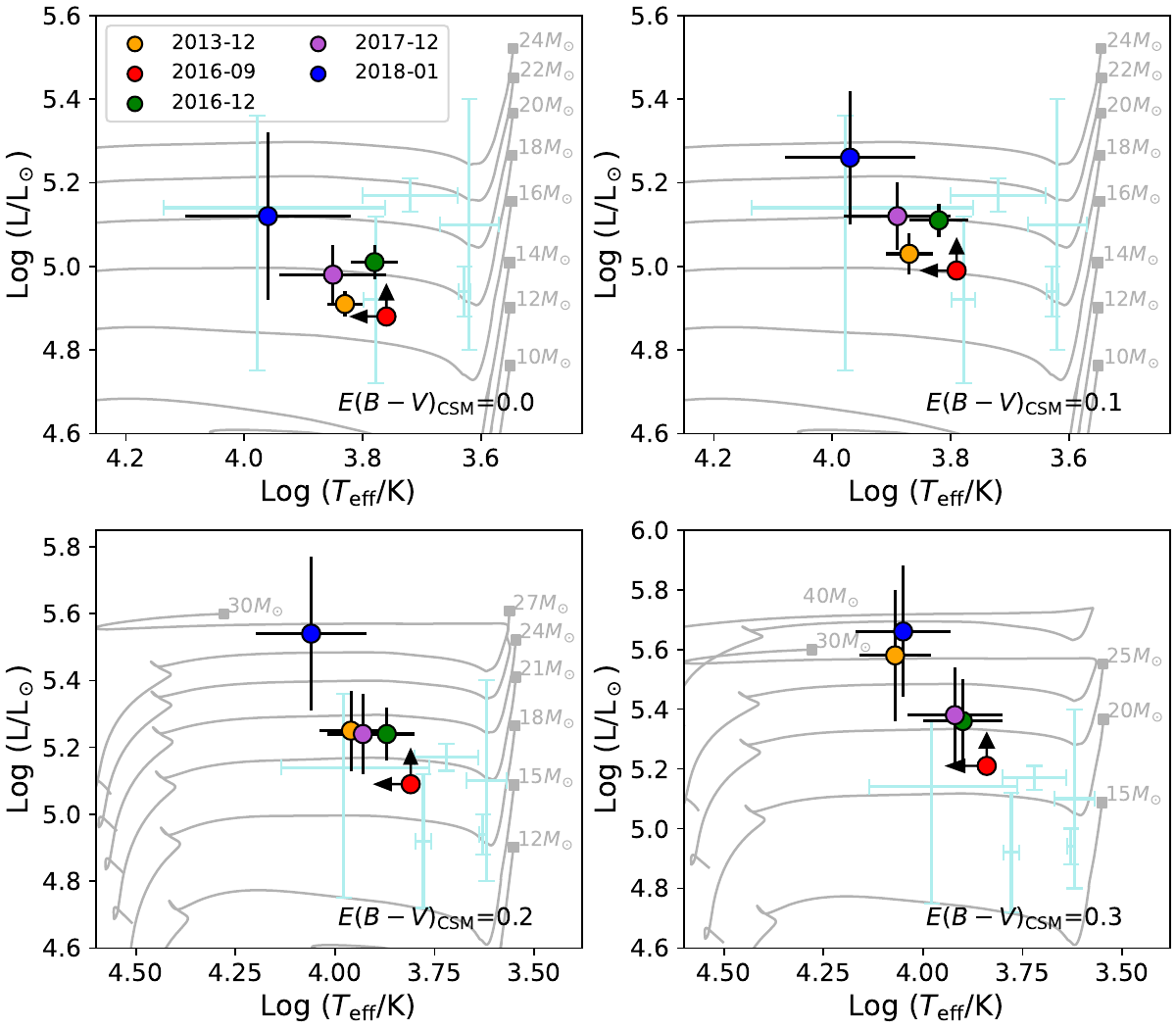}
    \caption{The evolution of positions on the HR diagram for the progenitor candidate, with different assumed circumstellar reddenings.
    Different epochs (yyyy-mm) are marked with different colors in the same way as in Figures~\ref{fig:SED}. 
    For reference, \textsc{bpass} single star evolution tracks (metallicity；$Z=0.01$) \citep{bpass.ref} are also plotted with different initial masses. Well-constrained progenitors of Type\,IIb SN\,1993J, SN\,2011dh, SN\,2013df, SN\,2016gkg, and SN\,2017gkk are plotted in lightblue.
    \label{fig:HRD}}
\end{figure*}

Figure~\ref{fig:HRD} illustrates the derived positions of the progenitor candidate on the Hertzsprung-Russell(HR) diagram at the 5 key epochs, along with the BPASS single-star evolutionary tracks \citep{bpass.ref} and other SNe~IIb progenitors with well-constrained parameters(SNe\,1993J, 2011dh, 2013df, 2016gkg, 2017gkk, \citealp{2004Maund,2011Maund,2014VanDyk,2022Kilpatrick,2017gkk.ref}).
The phenomenon mentioned before are all reflected in the HR diagram as the progenitor shows notable in both effective temperatures and bolometric luminosities.

When the circumstellar reddening is negligible(i.e. $E(B-V)_{\rm CSM}$\,=\,0.0~mag), the progenitor tend to be a yellow supergiant(YSG) star with log$(T_{\rm eff}/K)$ of 3.83--3.88 and and log($L/L_{\odot}$) of 4.92--5.03. It is difficult to estimate an accurate initial mass by comparing the variable luminosity with stellar evolutionary tracks in which variability is not involved. It seems that, however, the initial mass of the progenitor candidate is most likely to lie in range of 12--14~$M_{\odot}$. It is worth noting that, compared with a single star with the same initial mass, the partial stripping of the H-rich envelope by binary interaction only affects the effective temperature and leaves the core mass and luminosity almost unaffected \citep{2020Farrell, 2021Laplace}.

As the circumstellar reddening increases, the progenitor exhibits higher temperatures, larger luminosities, and consequently increased initial masses. The initial mass would lie in the range of 14--15~$M_{\odot}$ for $E(B-V)_{\rm CSM}$\,=\,0.1~mag and 15--18~$M_{\odot}$ for $E(B-V)_{\rm CSM}$\,=\,0.2~mag, if we exclude the epoch 2018 January during which the progenitor suddenly becomes overluminous(Fig.~\ref{fig:lc}). The inferred masses fall below the threshold, presumably $M_{\rm ini}\,>\,$25--30 $M_{\odot}$(e.g. \citealp{2001Massey,2005Meynet}), above which massive stars get stripped through their own stellar wind, thus supporting a binary origin of SNe~IIb \citep{1995Nomoto,2017Yoon,2024Dessart}. We deem even higher circumstellar reddening(e.g. $E(B-V)_{\rm CSM}$\,=\,0.3~mag) unlikely since the progenitor would become much hotter and more luminous than other SNe~IIb progenitors.

\begin{table}[htbp]
    \centering
    \caption{The derived stellar parameters for the progenitor candidate under different assumptions of reddening. \label{tab:sed}}
    \begin{tabular}{c|c|c|c}
\hline
\hline
Epoch & E(B-V)$_{\rm CSM}$ & log$(T_{\rm eff}/K)$ & log($L/L_{\odot}$) \\
(yyyy-mm) &  & & \\
\hline
2013-12 & 0.0 & 3.83(0.03) &  4.91(0.03) \\
2016-09 & 0.0 & $>$3.76(0.07) &  $>$4.88(0.09) \\
2016-12 & 0.0 & 3.78(0.04) &  5.01(0.04) \\
2017-12 & 0.0 & 3.85(0.09) &  4.98(0.07) \\
2018-01 & 0.0 & 3.96(0.14) &  5.12(0.20) \\
2013-12 & 0.1 & 3.87(0.04) &  5.03(0.05) \\
2016-09 & 0.1 & $>$3.79(0.09) &  $>$4.99(0.10) \\
2016-12 & 0.1 & 3.82(0.05) &  5.11(0.04) \\
2017-12 & 0.1 & 3.89(0.09) &  5.12(0.08) \\
2018-01 & 0.1 & 3.97(0.11) &  5.26(0.16) \\
2013-12 & 0.2 & 3.96(0.08) &  5.25(0.12) \\
2016-09 & 0.2 & $>$3.81(0.08) &  $>$5.09(0.08) \\
2016-12 & 0.2 & 3.87(0.07) &  5.24(0.08) \\
2017-12 & 0.2 & 3.93(0.09) &  5.24(0.12) \\
2018-01 & 0.2 & 4.06(0.14) &  5.54(0.23) \\
2013-12 & 0.3 & 4.07(0.09) &  5.58(0.22) \\
2016-09 & 0.3 & $>$3.84(0.08) &  $>$5.21(0.12) \\
2016-12 & 0.3 & 3.90(0.10) &  5.36(0.14) \\
2017-12 & 0.3 & 3.92(0.12) &  5.38(0.16) \\
2018-01 & 0.3 & 4.05(0.12) &  5.66(0.22) \\
\hline
    \end{tabular}
\end{table}

\section{Discussion and Conclusion}\label{sec:sum}

In this paper, we identify the progenitor star of the Type IIb SN\,2024abfo in the multi-epoch pre-explosion images from DECaLS in the $g$, $r$ and $z$ bands and from HST in the F814W band. By performing high-precision PSF photometry on single-epoch images, we detected significant and complex brightness and color variability in all filters.

A recent work also reported the progenitor candidate from HST/WFPC2 F814W-band, DECaLS $g,r,i,z$-band, and \textit{XMM-Newton}/OM $B,V$-band images \citep{Reguitti2025}. For DECaLS, they have used the cataloged average magnitudes over multiple epochs.
As demonstrated in Section~\ref{sec:decam}, the cataloged magnitudes are influenced by stellar variability and contamination from the nearby bright Source B.
Still, both works converge on an likely yellow supergiant progenitor for SN\,2024abfo.

Brightness variability has been widely observed for supergiant stars in the Local Universe, not pre-explosion but long before their core collapse.
Optical variability with amplitudes of about 0.01--1~mag on day to decade timescales has been observed for the substantial fraction of supergiant stars in the nearby galaxies \citep{2015Grammer,2018Conroy,2019Yang,2020Soraisam}. 
Similarly, long-term monitoring of luminous Galactic stars has revealed persistent photometric variability, alongside substantial color changes, indicative of effective temperature variations \citep{2019vanGenderen,2025vanGenderen}.

We find that the main cause of the variability is the stellar internal changes rather than external dust obscuration. 
The observed rapid changes with a $<$100-day timescale are difficult to reconcile with the quasi-static surface evolution predicted by standard one-dimensional stellar structure models, which assume hydrostatic equilibrium and typically yield luminosity variations on thermal timescales of $\sim10^4$ years or longer \citep{2013StellarStructureandEvolution}. 
It may not necessarily arise from global structural changes, but may reflect transient disturbances in the outer envelope, possibly driven by pulsations, episodic mass loss, and/or binary interactions \citep{1980Maeder,2006Kiss,2012Quataert,2014Shiode,2018Fuller,2024Schneider}.

Estimate of the progenitor's initial mass is complicated by the significant variability and the uncertain circumstellar extinction. The initial mass is likely to be 12--14, 14--15 and 15--18~$M_\odot$ for $E(B-V)_{\rm CSM}$\,=\,0, 0.1 and 0.2~mag, respectively.
We stress that an accurate estimate of the initial mass need further analysis of the light curves and nebular spectroscopy. Future late-time observations will also confirm whether this progenitor candidate has disappeared and reveal the putative binary companion that has survived the explosion.



This progenitor is the seventh directly detected SN~IIb progenitor. Thanks to the wealth of pre-explosion data，we had this first opportunity to observe the substantial brightness variability for a SN~IIb progenitor before explosion. 
Our findings not only extend the limited sample of SNe~IIb progenitors but also shed light on the final stages of massive star evolution preceding core-collapse.
Together with recent studies for progenitors of SNe~IIP \citep{2023Soraisam,2023Niu,Xiang2024,2019Rui}, the pre-explosion variability of supergiant stars seems to be quite common.
Therefore, this work also underscores the critical need for future sky surveys with multi-band, deep-detection capabilities, and long-baseline temporal coverage, such as Chinese Space Station Telescope \citep{2023Lin} and Legacy Survey of Space and Time \citep{2024Strotjohann,2024Petrecca}, to probe the complex dynamics of core-collapse SN progenitors. 

\section*{acknowledgments}

We thank the anonymous referee for helpful comments that improved our paper.

ZXN acknowledges support from the NSFC through Grant No. 12303039.
NCS’s research is funded by the NSFC Grant No. 12303051.
ZXN and NCS are also funded by the Strategic Priority Research Program of the Chinese Academy of Sciences with Grant No. XDB0550300 and the NSFC through Grant No. 12261141690.
This work is supported by the China Manned Space Program with grant No. CMS-CSST-2025-A14.
ZG is supported by the ANID FONDECYT Postdoctoral program No. 3220029. This work was funded by ANID, Millennium Science Initiative, AIM23-0001. 
This work was supported by CASSACA AND ANID through the China-Chile Joint Research Funding CCJRF2301.
WXL acknowledges the support from National Key R\&D Program of China (grant Nos. 2023YFA1607804, 2022YFA1602902, 2023YFA1607800, 2023YFA1608100), the National Natural Science Foundation of China(NSFC; grant Nos. 12120101003, 12373010, 12173051, and 12233008). WXL acknowledges the Strategic Priority Research Program of the Chinese Academy of Sciences with Grant Nos. XDB0550100 and XDB0550000. 
MS acknowledges support from the GBMF8477 grant(PI: Vicky Kalogera) and thanks Ping Chen and Chang Liu for helpful discussions. 
JFL acknowledges support from the NSFC through grants No. 11988101 and No. 11933004 and from the New Cornerstone Science Foundation through the New Cornerstone Investigator Program and the XPLORER PRIZE.

This research made use of Photutils, an Astropy package for
detection and photometry of astronomical sources \citep{Photutils.ref}.

The HST data presented in this article were obtained from the Mikulski Archive for Space Telescopes(MAST) at the Space Telescope Science Institute. The specific observations analyzed can be accessed via \dataset[doi: 10.17909/jvg3-1w15]{https://doi.org/10.17909/jvg3-1w15}.

This project used public archival data from the Dark Energy Survey(DES). Funding for the DES Projects has been provided by the U.S. Department of Energy, the U.S. National Science Foundation, the Ministry of Science and Education of Spain, the Science and Technology FacilitiesCouncil of the United Kingdom, the Higher Education Funding Council for England, the National Center for Supercomputing Applications at the University of Illinois at Urbana-Champaign, the Kavli Institute of Cosmological Physics at the University of Chicago, the Center for Cosmology and Astro-Particle Physics at the Ohio State University, the Mitchell Institute for Fundamental Physics and Astronomy at Texas A\&M University, Financiadora de Estudos e Projetos, Funda{\c c}{\~a}o Carlos Chagas Filho de Amparo {\`a} Pesquisa do Estado do Rio de Janeiro, Conselho Nacional de Desenvolvimento Cient{\'i}fico e Tecnol{\'o}gico and the Minist{\'e}rio da Ci{\^e}ncia, Tecnologia e Inova{\c c}{\~a}o, the Deutsche Forschungsgemeinschaft, and the Collaborating Institutions in the Dark Energy Survey.

The Collaborating Institutions are Argonne National Laboratory, the University of California at Santa Cruz, the University of Cambridge, Centro de Investigaciones Energ{\'e}ticas, Medioambientales y Tecnol{\'o}gicas-Madrid, the University of Chicago, University College London, the DES-Brazil Consortium, the University of Edinburgh, the Eidgen{\"o}ssische Technische Hochschule(ETH) Z{\"u}rich,  Fermi National Accelerator Laboratory, the University of Illinois at Urbana-Champaign, the Institut de Ci{\`e}ncies de l'Espai(IEEC/CSIC), the Institut de F{\'i}sica d'Altes Energies, Lawrence Berkeley National Laboratory, the Ludwig-Maximilians Universit{\"a}t M{\"u}nchen and the associated Excellence Cluster Universe, the University of Michigan, the National Optical Astronomy Observatory, the University of Nottingham, The Ohio State University, the OzDES Membership Consortium, the University of Pennsylvania, the University of Portsmouth, SLAC National Accelerator Laboratory, Stanford University, the University of Sussex, and Texas A\&M University.

Based in part on observations at Cerro Tololo Inter-American Observatory, National Optical Astronomy Observatory, which is operated by the Association of Universities for Research in Astronomy(AURA) under a cooperative agreement with the National Science Foundation.

\bibliography{sample631}{}
\bibliographystyle{aasjournal}


\end{CJK*}
\end{document}